\documentclass[preprint, 3p, twocolumn]{elsarticle}
\usepackage{graphicx, bm, epsfig, hyperref}
\biboptions{super, sort&compress}

\begin{document}
\title{Correspondences Between the Classical Electrostatic Thomson Problem and Atomic Electronic Structure}
\author{Tim LaFave Jr.}\corref{cor1}\fnref{fn1}
\cortext[cor1]{tjlafave@yahoo.com}
\fntext[fn1]{University of Texas at Dallas, Department of Electrical Engineering, Richardson, TX 75080}
\date{November 6, 2012}
\begin{abstract}
Correspondences between the Thomson Problem and atomic electron shell-filling patterns are observed as systematic non-uniformities in the distribution of potential energy necessary to change configurations of $N\le 100$ electrons into discrete geometries of neighboring $N\!-\!1$ systems. These non-uniformities yield electron energy pairs, intra-subshell pattern similarities with empirical ionization energy, and a salient pattern that coincides with size-normalized empirical ionization energies. Spatial symmetry limitations on discrete charges constrained to a spherical volume are conjectured as underlying physical mechanisms responsible for shell-filling patterns in atomic electronic structure and the Periodic Law.
\end{abstract}

\maketitle
\section{Introduction}
Quantum mechanical treatments of electrons in spherical quantum dots, or ``artificial atoms'',\cite{lafave-kastner1993} routinely exhibit correspondences to atomic-like shell-filling patterns by the appearance of abrupt jumps or dips in calculated energy or capacitance distributions as electrons are added to or removed from the system.\cite{lafave-bednarek1999, lafave-macucci1995, lafave-fujito1996, lafave-macucci1997, lafave-koskinen1997, lafave-lee1998, lafave-maksym2000, lafave-kouwenhoven2001, lafave-vorrath2003} Additionally, shell-filling is observed in ion trap models in which ions are subject to a spherical harmonic potential.\cite{lafave-rafac1991, lafave-wales1993, lafave-beekman1999, lafave-schiffer2003} An understanding of physical mechanisms responsible for shell-filling is useful to the engineering of tailorable electronic properties of quantum dots and ion traps as well as a better understanding of atomic electronic phenomena.

Electron shell-filling behavior has been observed in two-dimensional classical electrostatic models using a parabolic potential.\cite{lafave-bedanov1994} However, electrostatic treatments of three-dimensional artificial atoms have fallen short of yielding any observable shell-filling patterns.\cite{lafave-bednarek1999} Recently, similarities between classical electrostatic properties of spherical quantum dots and the distribution of empirical ionization energies of neutral atoms were reported for $N\le 32$ electrons\cite{lafave-lafave2006, lafave-lafave2008} when evaluated using the discrete charge dielectric model.\cite{lafave-lafave2011} The present paper builds on this previous work by identifying numerous correspondences between the electrostatic Thomson Problem of distributing equal point charges on a unit sphere and atomic electronic structure. 

Despite the diminished stature of J.J. Thomson's classical ``plum-pudding'' model\cite{lafave-thomson1904} among more accurate atomic models, the Thomson Problem has attracted considerable attention since the mid-twentieth century.\cite{lafave-whyte1952} The Thomson Problem has found use in practical applications including models of spherical viruses,\cite{lafave-caspar1962} fullerenes,\cite{lafave-garrido2000, lafave-kroto1985} drug encapsulant design,\cite{lafave-espinoza1999} and crystalline order on curved surfaces.\cite{lafave-bowick2002} Numerical solutions for many-$N$ electron systems have emerged in the last few decades using a variety of computational algorithms.\cite{lafave-erber1991, lafave-erber1995, lafave-edmundson1992, lafave-edmundson1993, lafave-altschuler1994, lafave-PGarrido1996, lafave-erber1997, lafave-garrido1999, lafave-altschuler2005, lafave-altschuler2006, lafave-wales2006} The Thomson Problem is now a benchmark for global optimization algorithms,\cite{lafave-altschuler2005, lafave-altschuler2006} yet its general solution remains an important unsolved mathematics problem.\cite{lafave-smale1998} 

A symmetry-dependent electrostatic potential energy distribution is obtained using numerical solutions of the Thomson Problem for $N\le 100$ electrons residing strictly on a unit sphere. This distribution exhibits many disparities (``jumps'' and ``dips'') that appear to be randomly distributed. However, upon closer inspection these disparities appear in a ``systematic''\cite{lafave-glasser1992} pattern shown here to be consistent with the pattern of atomic electron shell-filling as found in the form of the modern Periodic Table. A derivation of the symmetry-dependent potential energy distribution is given. A detailed description of its many correspondences with atomic electronic structure is provided in support of the conjecture that spatial symmetry limitations on discrete charges constrained to a spherical volume of space, as within a spherical dielectric or the central field of a nucleus, are underlying physical mechanisms responsible for electron shell-filling in quantum dots, ion traps, and atomic electronic structure. Additionally, a pattern of the largest energy disparities is shown to coincide with size-normalized empirical ionization energy data with discussion concerning relevant topological features of $N$-charge solutions and correspondence to shell-filling in atoms and ion traps. The systematic pattern of classical electrostatic symmetry-dependent energies consistent with atomic electron shell-filling is anticipated given the variety of neighboring geometric electron orbital shapes obtained from quantum mechanics ($s$, $p$, $d$, and $f$ orbitals).

For ease of verification, the reported results are based on data collected in an interactive database of numerical solutions of the Thomson Problem hosted by Syracuse University\cite{lafave-thomsonApplet} which may be compared with numerous other published sources.\cite{lafave-erber1991, lafave-erber1995, lafave-edmundson1992, lafave-edmundson1993, lafave-altschuler1994, lafave-PGarrido1996, lafave-erber1997, lafave-garrido1999, lafave-altschuler2005, lafave-altschuler2006, lafave-wales2006} 

\section{Discrete Symmetry Changes}
In the absence of a positively-charged spherical volume, electrons in the ``plum pudding'' model repel each other in such a manner that they naturally form solutions of the Thomson Problem.\cite{lafave-levin} These solutions are obtained by minimizing the total Coulomb repulsion energy

\begin{eqnarray}\label{eq:lafave-thomson}
U(N) = \sum_{i<j}^N \frac{1}{\left| r_i - r_j \right| }
\end{eqnarray}

\begin{figure}%
\begin{center}
\includegraphics[width=\columnwidth]{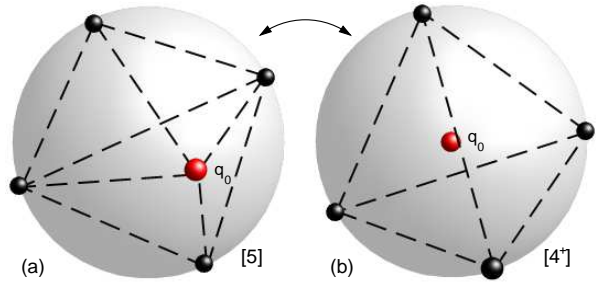}%
\end{center}
\caption{Discrete spatial symmetry changes. The 5-electron solution of the Thomson problem on (a) a unit sphere transforms into (b) the centered $[4^+]$ configuration having one charge, $q_0$, at the origin surrounded by the Thomson solution for 4-electrons on the unit sphere.}%
\label{fig:lafave-symmetricLoss}%
\end{figure}

{\noindent}of each $N$-electron system with $r_i$ and $r_j$ constrained to the surface of a unit sphere. An example of the $5$-electron solution is shown in Fig.~\ref{fig:lafave-symmetricLoss}a. The minimum energy is obtained with an electron at each ``pole'' of the unit sphere, and the remaining three electrons are located at vertices of an equilateral triangle about the ``equator''. Herein, the geometric configuration of each $N$-electron system is denoted in square brackets, $[N]$. 

To change the electrostatic electron configuration of a given $[N]$ solution of the Thomson Problem to the configuration of its neighboring $[N\!-\!1]$ solution such that the total number of electrons remains unchanged, a single electron is moved from the unit sphere to its origin. In general, the resulting electron distribution is one electron, $q_0$, at the origin, and the remaining $N\!-\!1$ electrons distributed on the unit sphere having the $[N\!-\!1]$ solution of the Thomson Problem. This centered configuration may be denoted by $[N\!-\!1^+]$, in which ``+'' indicates the presence of $q_0$. The total energy of any transformed system may be expressed as a function of $U([2])$, the energy of the two-electron solution of the Thomson Problem,

\begin{eqnarray}\label{eq:lafave-q0}
U([N\!-\!1^+]) = U([N\!-\!1])+ 2(N\!-\!1)U([2])
\end{eqnarray}

{\noindent}where the last term accounts for the interaction of $q_0$ with all $N\!-\!1$ electrons residing on the ``Thomson sphere''.

\begin{figure}%
\begin{center}
\includegraphics[width=0.8\columnwidth]{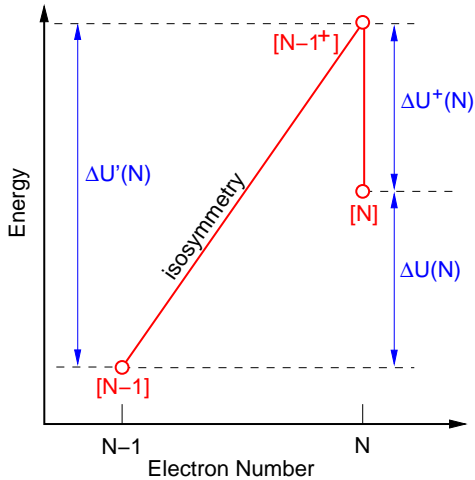}%
\end{center}
\caption{Discrete spatial symmetry changes in the Thomson problem. Changing the symmetry of a given $[N]$ configuration to the symmetry of its neighboring $[N\!-\!1]$ configuration while maintaining all $N$ electrons involves an intermediate $[N\!-\!1^+]$ centered configuration. This transition represents the symmetry-dependent component, $\Delta U^+(N)$, of the energy needed to remove a single electron. The remaining transition from $[N\!-\!1^+]$ to $[N\!-\!1]$ is the isosymmetric component whose energy, $\Delta U^{\prime}(N)$, is linearly dependent on $N$.}%
\label{fig:lafave-symmetryDiagram}%
\end{figure}

The $[4^+]$ solution shown schematically in Fig.~\ref{fig:lafave-symmetricLoss}b, consists of four electrons at vertices of a regular tetrahedron about $q_0$ at the origin. Symmetrically, the $[4^+]$ point group configuration is identical to the $[4]$ point group configuration\cite{lafave-wales1993} of the Thomson problem as $q_0$ interacts identically with all $N\!-\!1$ charges on the Thomson sphere. Using Eq.~\ref{eq:lafave-q0}, the energy difference as shown in Fig.~\ref{fig:lafave-symmetryDiagram}

\begin{eqnarray*}
\Delta U^{\prime}(N) &=& U([N\!-\!1^+]) - U([N\!-\!1])\\
&=& (N\!-\!1)
\end{eqnarray*}

{\noindent}is isosymmetric. Here, $U([2])=1/2$ for a unit sphere. In general, $U([N\!-\!1^+]) > U([N])$. Therefore, the energy difference $\Delta U(N)$ associated with the removal of an electron in the Thomson problem may be expressed,

\begin{eqnarray}\label{eq:lafave-UN}
\Delta U(N) &=& \Delta U^{\prime}(N) - \Delta U^+(N)\nonumber\\
&=& (N-1) - \Delta U^+(N)
\end{eqnarray}

{\noindent}as shown in Fig.~\ref{fig:lafave-symmetryDiagram}. Removal of an electron is dependent on a linear isosymmetric term in $N$ and a term resulting from a symmetry change between neighboring point groups. Energy differences due to this change in each $N$-electron system,

\begin{eqnarray*}
\Delta U^+(N) = U([N-1^+]) - U([N])
\end{eqnarray*}

\begin{figure*}%
\includegraphics[width=\textwidth]{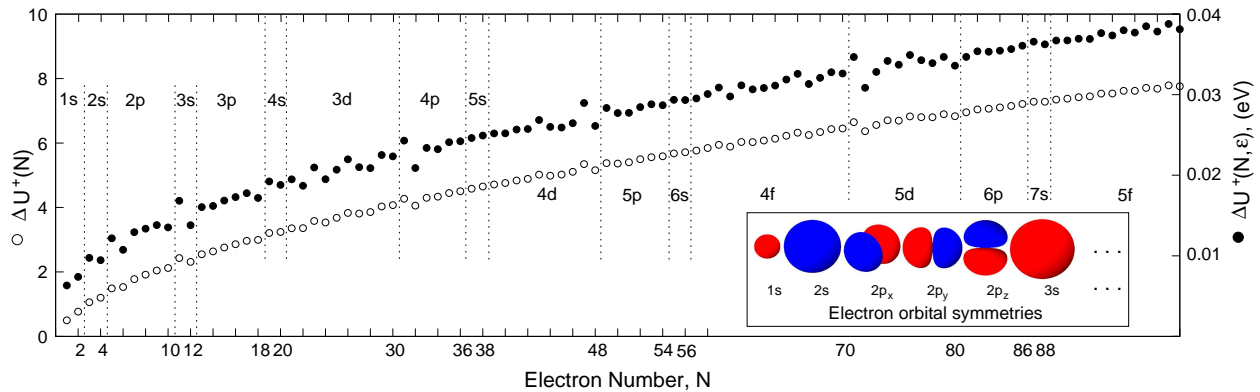}%
\caption{Correspondences between the Thomson problem and atomic electron shell-filling. Symmetry-dependent electrostatic energy, $\Delta U^+(N)$, associated transitions $[N]\to [N\!-\!1^+]$ are obtained by numerical solution of the Thomson Problem on a unit sphere in free space (open circles) and within a dielectric sphere (solid circles). Non-uniformities in this distribution correspond to changes in the spatial distribution of neighboring atomic orbitals (inset) as described in the text.}%
\label{fig:lafave-ThomsonDCD}%
\end{figure*}

{\noindent}are plotted in Fig.~\ref{fig:lafave-ThomsonDCD} (open circles) for $N\le 100$. 

\section{Correspondences with Atomic Electronic Structure}
Unlike classically-evaluated energies reported elsewhere,\cite{lafave-bednarek1999} the distribution of $\Delta U^{+}(N)$ is non-uniform. The energy differences are not equal, and their distribution is not ``smooth''. Importantly, however, this is a purely classical electrostatic distribution of energy differences associated exclusively with discrete geometric changes between [$N$] configurations and their neighboring [$N\!-\!1^+$] centered configurations. This $q_0$ transformation corresponds both with the single positive net-charge of a singly-ionized atom (charge sign symmetry) and the removal of the ionized electron through its mathematically-equivalent image charge, $q_0^{\prime}$, as may be realized within the context of a dielectric sphere. The energy distribution associated with this transformation, Fig.~\ref{fig:lafave-ThomsonDCD}, yields a systematic pattern of non-uniformities consistent with electron shell-filling patterns in atoms, exhibit distinct energy-pairs, feature intra-subshell energy patterns consistent with empirical intra-subshell ionization energies, and consists of a pattern of salient features consistent with size-normalized empirical ionization energies. Taken together, these many correspondences characterize Fig.~\ref{fig:lafave-ThomsonDCD} as an electrostatic ``fingerprint'' of the periodic table of elements resulting from a single ``shell'' of electrons in the Thomson Problem.

\subsection{Image Charges and Ionized Electrons}
Symmetry-dependent potential energy differences, $\Delta U^+(N,\varepsilon)$, of Thomson Problem solutions treated within a sphere of dielectric constant $\varepsilon=20\varepsilon_0$ and radius $a=100$nm, for illustrative purposes, are plotted in Fig.~\ref{fig:lafave-ThomsonDCD} (solid circles). This yields a similar energy distribution as before (open circles) but with much more pronounced non-uniformities owing partly to the minimization of the total stored energy with respect to the varying Thomson radius. Each charge, $q_i$, inside the dielectric sphere may be replaced by a mathematically equivalent image charge, $q_i^{\prime}$, {\it outside} the sphere. As $q_0$ approaches the origin, its image charge, $q_0^{\prime}$, moves away from the sphere. When $q_0$ arrives at the origin, $q_0^{\prime}$ is at such a distance that the remaining $N\!-\!1$ electrons reside in the [$N\!-\!1$] solution of the Thomson Problem. This argument is consistent with the isosymmetry between [$N\!-\!1^+$] and [$N\!-\!1$] shown in Fig.~\ref{fig:lafave-symmetryDiagram}. The physical interpretation of $q_0^{\prime}$ is the mathematical equivalent of the ionization of an electron from the system.

Charge sign symmetry is consistently involved in both the replacement of $q_0$ with its mathematically-equivalent image charge, $q_0^{\prime}$, as well as the appearance of a net positive charge located at the nucleus of a singly-ionized atom.

\subsection{Atomic Electron Shell-Filling Patterns}
Fig.~\ref{fig:lafave-ThomsonDCD} exhibits many non-uniform energy disparities (``jumps'' and ``dips'') which, as found in quantum mechanical evaluations of spherical quantum dots\cite{lafave-bednarek1999, lafave-macucci1995, lafave-fujito1996, lafave-macucci1997, lafave-koskinen1997, lafave-lee1998, lafave-maksym2000, lafave-kouwenhoven2001, lafave-vorrath2003} indicate divisions between electron subshells. The corresponding empirical atomic subshells are partitioned in Fig.~\ref{fig:lafave-ThomsonDCD} with dashed lines. Here, most subshells ``close'' with relatively lower energies than neighboring $N\!-\!1$ energies. Energies in Fig.~\ref{fig:lafave-ThomsonDCD} associated with $N$=4, 10, 12, 18, 20, 30, 36, 48, 54, 56, 70, 80, and 88 exhibit this trait. The few exceptions to this trend, $N$=2, 38, and 86, are indiscernible from the rising global trend of the distribution.

The ``opening'' of each subshell corresponds to a larger energy including large disparities at $N$= 3, 5, 11, 13, 19, 21, 31, 49, 55, 71, 81, 87, and 89 while others like $N$=37, 39, and 57 are less pronounced and indiscernible from the rising global trend. These disparities correspond well to the systematic distribution of electron subshells throughout the periodic table. 

The first disparity occurring between $N$=2 and 3 corresponds with the closing of the $1s$ shell in the periodic table and the opening of the $2s$ subshell. The second disparity occurring between $N$=4 and 5 corresponds with the closing of the spherical $2s$ subshell and the opening of the dumbbell-shaped $2p$ subshell. Likewise, the disparity between $N$=10 and 11 distinguishes the corresponding closing of the $2p$ subshell from the opening of the $3s$ subshell which is correspondingly closed with the disparity between $N$=12 and 13, opening the subsequent $3p$ subshell. In turn, the $3p$ subshell is correspondingly closed by a disparity between $N$=18 and 19, but the disparity between $N$=20 and 21 is smaller than all preceding disparities. This is not unexpected if the correspondence of the Thomson Problem with the periodic table is valid since the $3d$ energy level is known to be in close proximity to the $4s$ energy level. This is often cited as the underlying reason for half-filled low-lying $4s$ subshells in chromium ($Z$=24) and copper ($Z$=29). These are violations of common shell-filling rules.\cite{lafave-eisberg1974}

Lower energies in Fig.~\ref{fig:lafave-ThomsonDCD} correspond to more strongly-held electrons largely due to a higher degree of symmetry in the [$N$] configuration (low energy state) compared to the less energetically favorable symmetry of each neighboring [$N\!-\!1^{+}$] configuration (high energy state). Comparison with the distribution of empirical ionization energies of neutral atoms\cite{lafave-NIST} in Fig.~\ref{fig:lafave-ionization} demonstrates this correspondence. More energy is required to ionize an electron for values of $N$ that close electron subshells than those that open subsequent subshells. The symmetry-dependent energy distribution (Fig.~\ref{fig:lafave-ThomsonDCD}) is consistent with these empirical data.

\subsection{Electron Energy Pairs}
Consider carefully the distribution in Fig.~\ref{fig:lafave-ThomsonDCD} from left to right. Energies of neighboring $N$-electron systems tend to appear in pairs. For example, the pair $N$=(1,2) is well-isolated by a disparity from the energy at $N$=3 which forms another isolated pair with $N$=4. The $N$=(3,4) pair is well-isolated from the $N$=(5,6) pair which is in turn isolated from the energy at $N$=7. Though subtle, the $N$=(7,8) pair is isolated from the $N$=(9,10) energy pair, and the $N$=(11,12) pair is isolated from its neighbors but internally disparate. Similar pairings occur throughout the distribution in Fig.~\ref{fig:lafave-ThomsonDCD} with most pairs containing even-$N$ energies lower than their respective odd-$N$ energies. In a few instances the even-$N$ energy is higher such as in the pairs $N$=(1,2), (7,8), (15,16), (25,26), (35,36), (37,38), (45,46), and (51,52). In all, only 17 of 50 pairs shown in Fig.~\ref{fig:lafave-ThomsonDCD} exhibit lower odd-$N$ energies, and the majority of these seventeen are difficult to discern from the global trend of the distribution. Two such indiscernible examples, $N$=(37,38) and (85,86), close corresponding subshells.

\begin{figure*}%
\includegraphics[width=\textwidth]{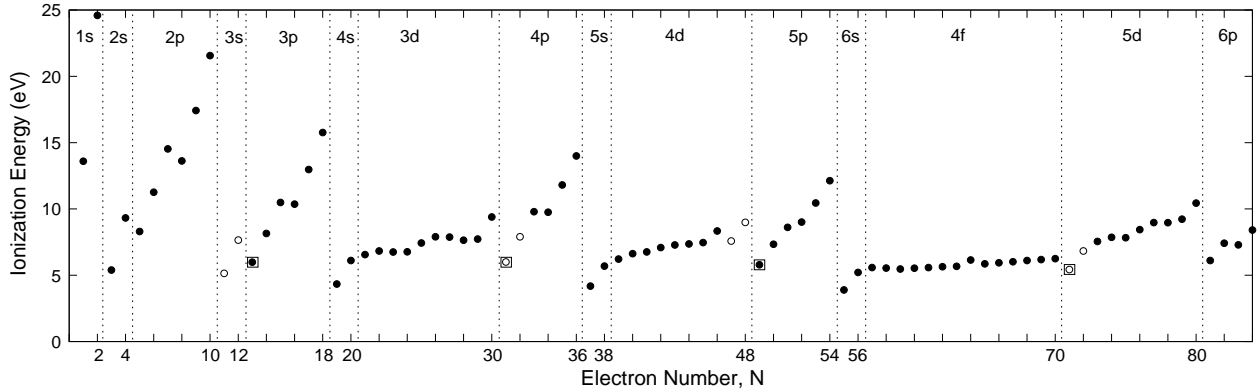}%
\caption{Empirical ionization energies of neutral atoms.\cite{lafave-NIST} For reference, ionization energies corresponding to energy pairs having largest internal disparity in $\Delta U^+(N)$ are shown as open circles. The lowest size-normalized empirical ionization locations are identified with squares.}%
\label{fig:lafave-ionization}%
\end{figure*}

Two characteristics of pairs are evident. Pairs may either ``rise'' (higher even-$N$ values) or ``fall'' (lower even-$N$ values), and pairs may be internally proximal or disparate. The most internally disparate pairs, $N$=(11,12), (31,32), (47,48) and (71,72) ``fall''. Notably, the odd-$N$ energies in these pairs rise significantly above the general trend of the distribution while their even-$N$ counterparts fall significantly below this trend. This suggests a strong preference for ``falling'' pairs. These four salient pairs are easily identified in the Thomson Problem treated in free space (open circles in Fig.~\ref{fig:lafave-ThomsonDCD}) and enunciated when treated in a dielectric sphere (solid circles).  The corresponding empirical ionization energies are shown as open circles in Fig.~\ref{fig:lafave-ionization} for reference.  Moreover, similar internally disparate pairs occur throughout the distribution of $N\le 500$-electron numerical solutions of the Thomson Problem as readily obtained using Eq.~\ref{eq:lafave-UN} and Ref.~\cite{lafave-thomsonApplet}. Notably, all are disparate falling pairs in glaring contrast to the rising global trend of the distribution.

The predominance of falling pairs reinforces a correspondence with the prevalence of closed electron subshells with even-$N$ systems in atomic electronic structure. On the other hand, rising pairs such as $N$=(25,26) and $(75, 76)$, clearly discernible from the global trend, occur at mid-points of the $3d$ and $5d$ subshells, respectively, and may therefore usefully guide future studies of the ``rule of stability'' concerning half-filled subshells. In contrast, $N=(43,44)$ is a falling pair, which may offer a clue to the instability of technetium ($Z=43$) if the present work is extended to nuclear structure.\cite{lafave-lafave2006}

\subsection{Intra-subshell Patterns}
As previously noted,\cite{lafave-lafave2008} the intra-subshell energy distribution between $21 \le N \le 30$ (Fig.~\ref{fig:lafave-ThomsonDCD}) exhibits a ``wavy'' behavior similar to the corresponding distribution of empirical ionization energies (Fig.~\ref{fig:lafave-ionization}) associated with the $3d$ atomic subshell. At the time of publication\cite{lafave-lafave2008} only evaluations $N\le 32$ in a dielectric sphere were reported. Here, electrostatic energies corresponding to $3d$, $4d$ and $5d$ subshells exhibit a similar ``wavy'' behavior with the notable exception of the first two pairs in the corresponding $4d$ subshell. Nonetheless, the distribution of empirical ionization data of neutral elements (Fig.~\ref{fig:lafave-ionization}) is consistent with this exception. This exception may be indicative of the fact that seven, $41\le Z\le 47$,\cite{lafave-lafave2006, lafave-sargentwelch} of the ten elements in the $4d$ subshell have empty or half-filled $5s$ subshells. With only two shell-filling violations in the $3d$ subshell and three in the $5d$ subshell ($Z$=77, 78, and 79) this is the largest number of violations in a single $d$ subshell. This ``wavy'' behavior may underlie the occurrence of known shell-filling rule violations.

Additionally, all corresponding $p$ subshells (Fig.~\ref{fig:lafave-ThomsonDCD}) are disparately isolated from their preceding electrostatic subshell energies with $6p$ being the least isolated from the $5d$ energies. The $2p$ and $3p$ subshells are well-isolated from their preceding $s$ subshells while the $4p$ subshell is well isolated from the $3d$ subshell. Notably, the $5p$ subshell is very well isolated from the $4d$ subshell. The $4d$ subshell closes with one of the four largest internally disparate energy pairs, again suggesting an underlying reason for the occurrence of so many shell-filling violations in the $4d$ subshell. Notably, corresponding $p$ subshell energy distributions all have only one or fewer ``rising'' pairs discernible from the global trend. Ionization energy trends within $p$ subshells are likewise similar. 

\subsection{Salient Correspondences}
\begin{figure*}%
\includegraphics[width=\textwidth]{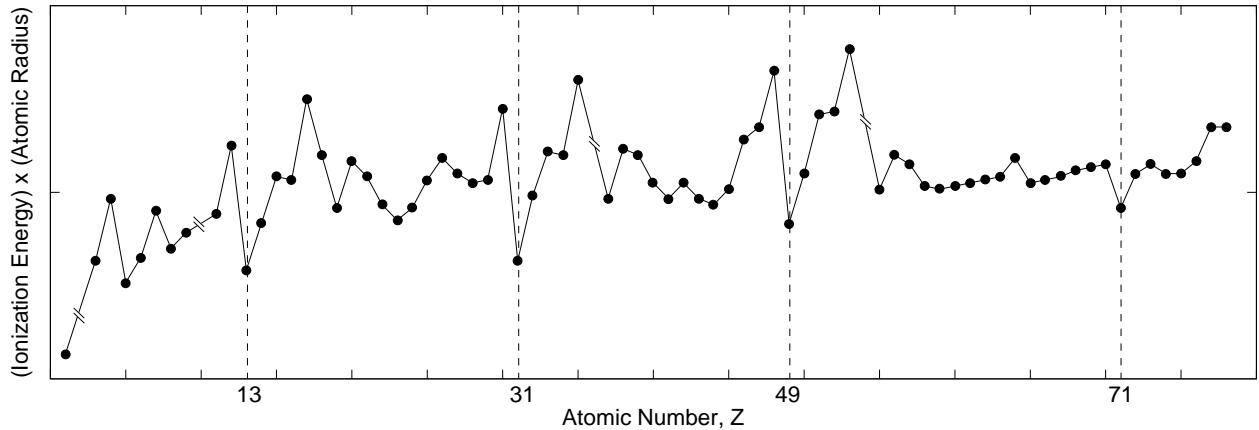}%
\caption{Empirical size-normalized ionization energies of neutral atoms.\cite{lafave-NIST, lafave-slater} The empirical distribution rises with increasing number of charges and includes four most disparate energy pairs, $Z=(12,13)$, $(30,31)$, $(48,49)$, and $(70,71) $.}%
\label{fig:lafave-empirical}%
\end{figure*}

Direct comparison between the Thomson Problem and atomic electronic structure is complicated as the former is constrained to a unit sphere while atoms vary in size. To account for this difference, the empirical ionization energy, $I$,  may be normalized with respect to size by empirical atomic radii, $R$.\cite{lafave-slater}. However, atomic radii are difficult to both define and measure, so only salient features of this approach are presently of interest. A plot of size-normalized empirical ionization data, $I\times R$, (Fig.~\ref{fig:lafave-empirical}) exhibits four widely disparate energy pairs with the four lowest even-$Z$ energies periodically spaced across a gradually increasing energy distribution. The four most-salient energy pairs appearing in the Thomson Problem, Fig.~\ref{fig:lafave-ThomsonDCD}, are in the immediate vicinity of these four disparate pairs in Fig.~\ref{fig:lafave-empirical}. As well, $\Delta U^+(N)$ and Fig.~\ref{fig:lafave-empirical} increase with respect to $N$ in a similar fashion. 

In order that the first and third salient symmetry-dependent energy pairs in Fig.~\ref{fig:lafave-ThomsonDCD} for $N=(11,12)$ and $(47,48)$ agree with the corresponding pairs in Fig.~\ref{fig:lafave-empirical}, $Z=(12,13)$ and $(48,49)$, they must shift to the right, while the other two pairs, $N=(31,32)$ and $(71,72)$ must shift to the left. To begin to understand this curious discrepancy, note that all four lower-$Z$ energies in the salient energy pairs in Fig.~\ref{fig:lafave-empirical} correspond to ionization energies that open $p-$ and $d-$ subshells (represented by squares in Fig.~\ref{fig:lafave-ionization}). These ionization energies form a set of secondary lowest energies in Fig.~\ref{fig:lafave-ionization}. Effectively, the primary set of lowest ionization energies that open $s$-shells (begin each period on the periodic table) are size-dependent, whereas the ionization energies that open secondary subshells are symmetry-dependent. Note that the size-normalization process is isosymmetric and removes {\it most} of the size-dependence in the distribution of ionization energies. Indeed, the size-dependent lowest-energies in Fig.~\ref{fig:lafave-ionization} do not stand out in the symmetry-dependent distribution of size-normalized empirical ionization energy (Fig.~\ref{fig:lafave-empirical}). 

Differences between Figs.~\ref{fig:lafave-ThomsonDCD} and \ref{fig:lafave-empirical} must be related to symmetry properties of discrete charges. The fact that Fig.~\ref{fig:lafave-ThomsonDCD} disparate energy pairs must shift either left or right ($N\pm 1$) is clear since lower symmetry-dependent $\Delta U^+(N)$ correspond to more tightly-bound electrons that require more energy to ionize than electron configurations having lesser degrees of spatial symmetry.

Topologically, Thomson Problem solutions of $N=12$ and $48$ share features while $N=32$ and $72$ share other features. The latter two solutions may be generated by their number of vertices, $V=10(h^2+k^2+hk)+2$, ($h \ge k > 0$ integers)\cite{lafave-altschuler1997, lafave-PGarrido1997, lafave-wales2009} obtained by consideration of the Euler Formula for convex polyhedra having triangular faces. The special case of $(h=1,k=0)$ yields the icosohedral $N=12$ solution, with no other integer value of $h$ with $k=0$ yielding Thomson Problem solutions. In the former solutions, $N=12$ and $48$ may be obtained by $V_5 + 2V_4 - V_7 = 12+12Q$ for polyhedra having vertices that are pentamers ($V_5$), tetramers ($V_4$), and heptamers ($V_7$) with the remaining vertices as hexamers.\cite{lafave-altschuler1997} The $N=48$ solution includes 6 quadrilateral faces (24 pentamers and 24 hexamers), and $N=12$ is again a special case for $Q=0$ (12 pentamers). Hence, the $N=(11,12)$ and $(47,48)$ pairs topologically differ from $N=(31,32)$ and $(71,72)$ and are expected to correspond differently to empirical features.

The character of each pair of disparities is also intimately linked to shell-filling. Both $N=12$ and $48$ close the respective subshell (Fig.~\ref{fig:lafave-ionization}), however, neither $N=32$ nor $72$ close a subshell. Instead, the associated odd-$N$ energies open subshells as does the unique topologically shared case of $N=(11,12)$. To further substantiate this observation, consider correspondences with ion trap models in which multiple shells are obtained. The $N=12$ and $48$ solutions of the Thomson Problem correspond to shell-closings in ion traps, noting that for $2\le N \le 12$ a single ion shell is obtained with minimum global energy and a second shell appears in the range $13 \le N \le 60$ ions with $48$ ions in the outer shell for $N=60$.\cite{lafave-rafac1991, lafave-wales1993, lafave-beekman1999, lafave-schiffer2003} In ion trap models, $N=32$ and $N=72$ do not exhibit shell-closing features, consistent with Fig.~\ref{fig:lafave-ThomsonDCD}. Subsequently, the electrostatic energy pairs $N=(11,12)$ and $(47,48)$ may be associated with shell-closing due to their lower even-$N$ energies while $N=(31,32)$ and $(71,72)$ may be considered as associated with shell-opening with their higher odd-$N$ energies. Consequently, it appears the necessary shifts between Figs.~\ref{fig:lafave-ThomsonDCD} and \ref{fig:lafave-empirical} are in the direction of the nearest shell boundary. In the case of $N=12$, given its high degree of symmetry and corresponding shell-closing in ion trap models, there is a preference to shift to the right. As well, the disparity related to $N=48$ may be indicative of the many shell-filling rule violations in the preceding 4$d$ subshell as noted earlier.

\section{Discussion \& Conclusions}
Identification of several correspondences between electrostatic solutions of the Thomson Problem and shell-filling patterns in natural atomic electronic structure originated in classical studies of spherical quantum dots.\cite{lafave-lafave2006} The classical symmetry-dependent electrostatic potential energy distribution, $\Delta U^{+}(N)$, of Fig.~\ref{fig:lafave-ThomsonDCD}, however, stems from the more fundamental Thomson Problem. Underlying all correspondences with atomic electron shell-filling are spatial symmetry limitations imposed on each $N$-electron system. For example, electrostatic energies at $N$=4 and $5$ in Fig.~\ref{fig:lafave-ThomsonDCD} are correspondingly quite disparate, as one might expect given a difference in the geometric shapes of the neighboring spherical $2s$ and dumbbell-shaped $2p$ orbitals. The electrostatic configuration of $[4]$ includes electrons at vertices of a regular tetrahedron in which all electrons are equidistant. This symmetry property, which permits all electrons in $[4]$ to have equal energy, is unavailable to any system of 5 electrons in three-dimensional space. Indeed, global equidistance is impossible for all $N\ge 5$ systems. Physical impossibilities such as this restrict the number of possible electrostatic configurations of discrete charges and related physical quantities. However, equal energy distribution is not exclusively obtained by global equidistance among electrons. Several other Thomson Problem solutions have all electrons residing in the same energy level, $N=6$, $8$, $12$ and $24$, in which other symmetry operations are involved.

The conjecture of spatial limitations on discrete charges as a physical mechanism responsible for electron shell-filling is supported by the many correspondences reported here between the classical electrostatic Thomson Problem and naturally-occurring atomic electronic structure. The $q_0$ transformation of a single charge to the origin of the Thomson sphere and the mathematically equivalent loss of its image charge, $q_0^{\prime}$, supports the correspondence of $\Delta U^{+}(N)$ to the ionization energy associated with a singly-ionized atom. However, charges in the Thomson Problem are constrained to a fixed unit sphere. Insertion of the Thomson Problem into a dielectric sphere yields more pronounced non-uniformities in the distribution of $\Delta U^+(N)$ in Fig.~\ref{fig:lafave-ThomsonDCD} as the unit Thomson radius constraint is relaxed to minimize the sum of all electrostatic interaction terms within the sphere.\cite{lafave-lafave2011} This suggests that a screening parameter is important to future developments as it is involved in the spherical jellium model\cite{lafave-ekardt1984} which presumes a ``uniform'' positive background charge within a spherical volume (cf.~the ``plum pudding'' model) to neutralize the total charge of the system.

Further, relaxing the constraint of a common radius among all electrons within the dielectric sphere model, beginning with $N=5$, lower energies are obtained when the ``polar'' radius is varied with respect to the ``equatorial'' radius (cf. Fig.~\ref{fig:lafave-symmetricLoss}a). As the dielectric constant of the sphere increases the polar radius is greater than the equatorial radius until roughly $3.7\le \varepsilon/\varepsilon_0 \le 4.0$, after which the equatorial radius exceeds the polar radius for large dielectric constants. This tendency toward ellipsoidal configurations is consistent with ellipsoidal equilibrium results of free-electron metal clusters.\cite{lafave-clemenger1985} 

Non-uniformities in the energy distribution in Fig.~\ref{fig:lafave-ThomsonDCD} are uncharacteristic of classically-derived energy distributions which suggest no shell-filling behavior.\cite{lafave-bednarek1999} However, similar to the correlation of non-uniformities arising in quantum mechanical evaluations of spherical quantum dots, non-uniformities appearing in the present classical electrostatic evaluation have been shown to correspond exceedingly well to electron shell-filling patterns in natural atomic electronic structure. In this regard, Fig.~\ref{fig:lafave-ThomsonDCD} is a classical electrostatic ``fingerprint'' of the periodic table. Included in this ``fingerprint'' are intra-subshell energy patterns consistent with empirical ionization energy patterns and energy disparities that coincide with electron shell boundaries with remarkable fidelity. In addition to these correspondences is a secondary fingerprint. The four most internally-disparate energy pairs occur in the immediate vicinity of corresponding size-normalized empirical ionization energies (Fig.~\ref{fig:lafave-empirical}), and a similar rise in energy is observed as with the empirical distribution. Shared features among these disparate pairs are worth further investigation as they correspond intimately to subshell-filling phenomena through spatial symmetry. The intertwined symmetry and isosymmetry (size) components of ionization energy may be usefully explored within the context of the principle of maximum symmetry in which energy obtained by maximally symmetric charge configurations may yield particularly low and high energy {\it minima}\cite{lafave-wales1998, lafave-wales1998a} as observed in the primary and secondary sets of ionization minima in Fig.~\ref{fig:lafave-ionization} as contrasted with the size-normalized empirical ionization energies in Fig.~\ref{fig:lafave-empirical}. This is intimately connected to a better comprehension of the physical mechanisms underlying the Periodic Law of elements.

Missing from the evaluation of size-normalized ionization energies in Fig.~\ref{fig:lafave-empirical} is further partitioning into isosymmetric (size) and symmetry-dependent components (cf.~Fig.~\ref{fig:lafave-symmetryDiagram}) of the ionization process. In particular, along one ionization ``path'', (cf.~Fig.~\ref{fig:lafave-symmetryDiagram}), at some distant point in its trajectory away from the system the ionized electron (equivalent to $q_0^{\prime}$) still interacts with the remaining $N\!-\!1$ electrons such that they are in their equilibrium $[N\!-\!1^+]$ configuration. From this point to full removal of the electron the $N\!-\!1$ electron system equilibrates to the final size of the resulting ion. Along another (likely, more realistic) path, the isosymmetric size reduction component of ionization energy exhibits {\it while} the ionized electron exits the system (a direct path between $[N]$ and $[N\!-\!1]$ in Fig.~\ref{fig:lafave-symmetryDiagram}). Both energy-paths yield the same result. However, appropriate partitioning for direct comparison with the Thomson Problem requires an evaluation of $[N\!-\!1^+]$ in empirical size-normalized atomic systems, which is complicated by the fact that not all electrons reside at the same Thomson radius. As a result, the empirical isosymmetric term is {\it not} expected to be linear with respect to $N$.


\begin{thebibliography}{100}
\bibitem{lafave-kastner1993}
M. A. Kastner, Physics Today {\bf 46}, 24 (1993).

\bibitem{lafave-bednarek1999}
S. Bednarek, B. Szafran, and J. Adamowski, Phys. Rev. B. {\bf 59}, 13036 (1999).

\bibitem{lafave-macucci1995}
M. Macucci, K. Hess, and G. J. Iafrate, J. Appl. Phys. {\bf 77}, 3267 (1995). 

\bibitem{lafave-fujito1996}
M. Fujito, A. Natori, and H. Yasunaga, Phys. Rev. B {\bf 53}, 9952 (1996).

\bibitem{lafave-macucci1997}
M. Macucci, K. Hess, and G. J. Iafrate, Phys. Rev. B {\bf 55}, R4879 (1997).

\bibitem{lafave-koskinen1997}
M. Koskinen, M. Manninen, and S. M. Reimann, Phys. Rev. Lett. {\bf 79}, 1389 (1997).

\bibitem{lafave-lee1998}
I.-H. Lee, V. Rao, R. M. Martin, and J.-P. Leburton, Phys. Rev. B {\bf 57,} 9035 (1998).

\bibitem{lafave-maksym2000}
P. A. Maksym, H. Imamura, G. P. Mallon, and H. Aoki, J. Phys.: Condens. Matter {\bf 12}, R299 (2000).

\bibitem{lafave-kouwenhoven2001}
L. P. Kouwenhoven, D. G. Austing, and S. Tarucha, Rep. Prog. Phys. {\bf 64}, 701 (2001).

\bibitem{lafave-vorrath2003}
T. Vorrath and R. Bl\"umel, Euro. Phys. J. B {\bf 32}, 227 (2003).

\bibitem{lafave-rafac1991}
R. Rafac, J. P. Schiffer, J. S. Hangst, D. H. Dubin, and D. J. Wales, Proc. Natl. Acad. Sci., {\bf 88} 483-486 (1991).

\bibitem{lafave-wales1993}
D. J. Wales and A. M. Lee, Phys. Rev. A, {\bf 47}(1) 380-392 (1993).

\bibitem{lafave-beekman1999}
R. A. Beekman, M. R. Roussel, and P. J. Wilson, Phys. Rev. A, {\bf 59}(1) 503-511 (1999).

\bibitem{lafave-schiffer2003}
J. P. Schiffer, J. Phys. B: Mol. Opt. Phys {\bf 36} 511-523 (2003).

\bibitem{lafave-bedanov1994}
V. M. Bedanov and F. M. Peeters, Phys. Rev. B {\bf 49}(4) 2667-2676 (1994).

\bibitem{lafave-lafave2006}
T. LaFave Jr, Ph.D. dissertation, University of North Carolina, Charlotte, 2006.

\bibitem{lafave-lafave2008}
T. LaFave Jr. and R. Tsu, Microelectron. J. {\bf 39}, 617 (2008).

\bibitem{lafave-lafave2011}
T. LaFave Jr., J. Electrostatics {\bf 69}, 414 (2011).

\bibitem{lafave-thomson1904}
J. J. Thomson, Philos. Mag. {\bf 7}, 237 (1904).

\bibitem{lafave-whyte1952}
L. L. Whyte, Am. Math. Month. {\bf 59}, 606 (1952).

\bibitem{lafave-caspar1962}
D. L. D. Caspar and A. Klug, Cold Spring Harbor Symp. on Quant. Biology {\bf 27}, 1 (1962).

\bibitem{lafave-garrido2000}
A. P\'erez-Garrido, Phys. Rev. B {\bf 62}, 6979 (2000).

\bibitem{lafave-kroto1985}
H. W. Kroto, J. R. Heath, S. C. O'Brien, R. F. Curl, and R. E. Smalley, Nature {\bf 318}, 162 (1985).

\bibitem{lafave-espinoza1999}
H. Aranda-Espinoza, Y. Chen, N. Dan, T. C. Lubensky, P. Nelson, L. Ramos, and D. A. Weitz, Science {\bf 285}, 394 (1999).

\bibitem{lafave-bowick2002}
M. Bowick, A. Cacciuto, D. R. Nelson, and A. Travesset, Phys. Rev. Lett. {\bf 89}, 185502 (2002).

\bibitem{lafave-erber1991}
T. Erber and G. M. Hockney, J. Phys. A: Math. Gen. {\bf 24}, L1369 (1991).

\bibitem{lafave-erber1995}
T. Erber and G. M. Hockney, Phys. Rev Lett. {\bf 74}, 1482 (1995).

\bibitem{lafave-edmundson1992}
J. R. Edmundson, Acta Cryst. A {\bf 48}, 60 (1992).

\bibitem{lafave-edmundson1993}
J. R. Edmundson, Acta Cryst. A {\bf 49}, 648 (1993).

\bibitem{lafave-altschuler1994}
E. L. Altschuler, T. J. Williams, E. R. Ratner, F. Dowla, and F. Wooten, Phys. Rev. Lett. {\bf 72}, 2671 (1994).

\bibitem{lafave-PGarrido1996}
A. P\'erez-Garrido, M. Ortu\~no, E. Cuevas, and J. Ruiz, J. Phys. A: Math. Gen. {\bf 29}, 1973 (1996).

\bibitem{lafave-erber1997}
T. Erber and G. M. Hockney, Adv. Chem. Phys. {\bf 98}, 495 (1997).

\bibitem{lafave-garrido1999}
A. P\'erez-Garrido and M. A. Moore, Phys. Rev. B {\bf 60}, 15628 (1999).

\bibitem{lafave-altschuler2005}
E. L. Altschuler and A. P\'erez-Garrido, Phys. Rev. E {\bf 71}, 047703 (2005).

\bibitem{lafave-altschuler2006}
E. L. Altschuler and A. P\'erez-Garrido, Phys. Rev. E {\bf 73}, 036108 (2006).

\bibitem{lafave-wales2006}
D. J. Wales and S. Ulker, Phys. Rev. B {\bf 74}, 212101 (2006).

\bibitem{lafave-smale1998}
S. Smale, Math. Intell. {\bf 20}, 7 (1998).

\bibitem{lafave-glasser1992}
L. Glasser and A. G. Every, J. Phys. A: Math. Gen {\bf 25} 2473-2482 (1992).

\bibitem{lafave-thomsonApplet}
Thomson Applet, \url{http://thomson.phy.syr.edu}

\bibitem{lafave-levin}
Y. Levin and J. J. Arenzon, Europhys. Lett. {\bf 63}, 415 (2003).

\bibitem{lafave-eisberg1974}
R. Eisberg and R. Resnick, {\it Quantum Physics of Atoms, Molecules, Solids, Nuclei, and Particles} (John Wiley \& Sons, 1974) pp.~357-365.

\bibitem{lafave-NIST}
NIST: Ground levels and ionization energies for the neutral atoms \url{http://physics.nist.gov/PhysRefData}

\bibitem{lafave-sargentwelch}
Table of Periodic Properties of the Elements, Sargent-Welch, 1980.

\bibitem{lafave-slater}
J. C. Slater, J. Chem. Phys. {\bf 41}(10) 3199 (1964).

\bibitem{lafave-altschuler1997}
E. L. Altschuler, T. J. Williams, E. R. Ratner, R. Tipton, R. Stong, F. Dowla, and F. Wooten, Phys. Rev. Lett. {\bf 78}(14) 2681-2685 (1997).

\bibitem{lafave-PGarrido1997}
A. P\'erez-Garrido, M. J. W. Dodgson, M. A. Moore, M. Ortu\~{n}o, and A. D\'iaz-S\'anchez, Phys. Rev. Lett., {\bf 79}(7) 1417 (1997).

\bibitem{lafave-wales2009}
D. J. Wales, H. McKay, and E. L. Altschuler, Phys. Rev. B, {\bf 79} 224115 (2009).

\bibitem{lafave-ekardt1984}
W. Ekardt, Phys. Rev. B, {\bf 39} 1558-1564 (1984).

\bibitem{lafave-clemenger1985}
K. Clemenger, Phys. Rev. B, {\bf 32} 1359-1362 (1985).

\bibitem{lafave-wales1998}
D. J. Wales, Chem. Phys. Lett., {\bf 285} 330-336 (1998).

\bibitem{lafave-wales1998a}
D. J. Wales, Chem. Phys. Lett., {\bf 294} 262 (1998).

\end{thebibliography}
\end{document}